\documentclass{aastex} 
\usepackage{emulateapj5}
\usepackage{natbib}

\shorttitle{Co-orbital Vortices of a Protoplanet}

\shortauthors{Koller, Li, \& Lin}

\begin{document}

\title{Vortices in the Co-orbital Region of an Embedded
Protoplanet}

\author{Josef Koller\altaffilmark{1,2}, Hui Li\altaffilmark{1},
and Douglas N.C. Lin \altaffilmark{3}}
\altaffiltext{1}{
Applied Physics Division, X-1, MS P225, Los Alamos National
Laboratory, NM 87545; jkoller@lanl.gov; hli@lanl.gov}
\altaffiltext{2}{
Department of Physics \& Astronomy, MS-108, Rice
University, Houston, TX 77005}
\altaffiltext{3}{UCO/Lick Observatory, University of California,
Santa Cruz, CA 95064; lin@ucolick.org}

\begin{abstract}
We present global 2-D inviscid disk simulations with an embedded 
planet, emphasizing the non-linear dynamics in its co-orbital 
region. We find that the potential vorticity of the flow in this
region is not conserved due to the presence of two spiral shocks
produced by the planet. As the system evolves, the potential vorticity 
profile develops extrema (inflection points) which eventually
render the flow unstable. Vortices are produced in association with
the potential vorticity minima. Born in the separatrix region,
these vortices experience close-encounters with the planet, 
consequently exerting strong torques on the planet. The existence
of these vortices have important implications on understanding
the migration rates of low mass planets.
\end{abstract}


\sloppy
\section{INTRODUCTION}

The discovery of a large number of short-period extrasolar planets  
has generated renewed interests in the study of
tidally induced migration of protoplanets in disks.  The basis of
migration is the gravitational interaction between a gaseous disk and
an orbiting perturber, and has been analyzed well before extrasolar
planets were actually discovered
\citep{GoldreichTremaine79,GoldreichTremaine80,LinPap86a,LinPap86b}. It
has been suggested that this tidal interaction induces the embedded
protoplanet to migrate under two different circumstances.  In Type~I
migration, the protoplanet mass is small and the disk structure is
weakly perturbed.  In this limit, the disk response can be analyzed
with a quasi-linear approximation.  Linear calculations have revealed
an intrinsic imbalance of inner and outer disk torques \citep{Ward97},
so the planet could migrate quite quickly inward.  In Type II
migration, the protoplanet mass is sufficiently large for it to open
up a gap in the disk \citep{PapLin84}. The protoplanet's orbital
evolution is locked to the evolution of the disk, which is likely to
be on the viscous timescale \citep{NelsonPapMassetKley00}. Both types
of migration have been basically confirmed by the non-linear numerical
simulations.

The high mobility of small mass protoplanets or embryo cores of giant
planets from Type I migration has presented a challenge to the
formation of giant planets.  The absence of deep oceans is contrary to
the inference that any terrestrial planet in the solar system may be
formed well outside the snow-line and migrated to their present
location.  It is thus imperative to explore the cause which may have
inhibited the Type I migration.  One potential contributor to this
process is the disk response to low mass planets, especially in the
co-orbital region where only sketchy analytical arguments have been
developed. There are, however, two notable recent exceptions. Masset
(2001, 2002) has evaluated the co-orbital corotation torque and found
that they can play a crucial role for the migration rate. Extensive
numerical simulations of viscous disks have been performed to
understand the torque dependence on the planet mass, sound speed, disk
viscosity, and the torque saturation mechanism. One generic behavior
he found is that the total torque on the planet first shows an
oscillation consistent with the librating fluid motion in the
co-orbital region then it levels off to a constant due to the disk
viscosity.

Another study is by Balmforth \& Korycansky (2001), who used matched
asymptotic expansions to explore the non-linear dynamics of the flows
in the co-orbital region.  They investigated the role of re-arranging
the potential vorticity profile as a response to the planet's driving
and its role in the corotation torque saturation. For inviscid flows,
they also noticed the formation of vortices which they attributed to
the secondary instabilities in a forced critical layer problem (the
critical layer in this case is the corotation resonance).

In this {\em Letter}, we present results on the non-linear dynamics 
of the co-orbital region, based on direct, high resolution
numerical simulations (\S 2). Our approach is quite 
similar to that of Masset's except that we have focused on the 
inviscid limit simulations. As a consequence, the total torque on 
the planet exhibits some interesting 
new behavior, instead of leveling off to a constant as in the case 
of viscous disks. 
Some of the findings are similar to what Balmforth \& 
Korycansky have found in the excitation of vortices (\S 3). 
We provide an interpretation of why these vortices are produced in
\S 4 and discuss the implications of our results in \S 5.


\section{Initial Set-up}

We assume that the protoplanetary disk is thin and can be described
by the inviscid 2-D Euler equations in a cylindrical $(r, \phi)$ plane
with vertically integrated quantities.  Two gravitating bodies, a
central star and a protoplanet, are located at $r=0$ and $r=1$
respectively and the 2-D disk is modeled between $0.4 \leq r \leq 2$.
The planet is on a fixed circular orbit at $r=1$.  The self-gravity of
the disk is not included.

The mass ratio between the planet and the central star is taken to be
$\mu=M_{p}/M_{\ast}=10^{-4}$. Its Hill (Roche) radius is $r_H =
(\mu/3)^{1/3} = 0.032$. The disk is assumed to be isothermal with a
constant temperature throughout the simulation region (i.e., it is
attached to a thermal bath). The isothermal sound speed, scaled by the
Keplerian rotation speed $v_\phi$ at $r=1$ is $c_s/v_{\phi,r=1}=H/r$
where $H$ is the disk scale height.  We used a set of three sound
speeds $c_s=0.04, 0.05, 0.06$.  In the simulations presented here, we
choose an initial surface density profile with $\Sigma(r) \propto
r^{-3/2}$, in order that the ratio of vorticity to surface density
(potential vorticity) has a flat radial
profile $\zeta = (\nabla\times {\bf v})_z/\Sigma \approx const.$ (Note
that $\zeta$ has a small deviation from being a precise constant due
to the finite pressure gradient which slightly modifies $v_\phi$ from
its Keplerian value.)  With this choice of initial condition, we avoid
the generation of inflection points due to the rearrangement of
$\zeta$ distribution as it is carried by the stream lines.  We have
also made other runs which have an initial potential vorticity
gradient and we find that the results presented here also hold for
them.  The detailed analysis of these results will be presented
elsewhere.

Since $r_H < H$ for all three different sound speeds, the planet is
embedded. The planet's gravitational potential is softened by an
approximate 3-D treatment.
The key approach is to take the surface density 
at each cell spreading it out vertically based on a Gaussian density
profile with a scale height determined by the local temperature.
The gravitional force between the planet and the disk matter is 
calculated using this 3-D configuration but all the forces with
the same $(r, \phi)$ positions are summed together. This
gives a correction factor $C(r,\phi)$ to the gravitational forces,
which effectively softens the gravitational potential. The torque
overestimation in a 2-D geometry was studied in
\cite{DAngeloKleyHenning03} which arises only from locations near the
planet. As a test if our key finding is affected by the 3-D treatment,
we excluded several Roche radii from the torque calculation and show
that it is not the case.
 
Although there are some accumulation of gas near the
planet, we do not allow any disk gas to be accreted onto it.  This
boundary condition is equivalent to that of sufficiently low-mass
protoplanets with quasi hydrostatic envelopes (Pollack {\it et al.}
1996).  The planet's potential is switched on over 10 orbits in order
for the disk to make an adiabatic adjustment. (We have used
other, longer turn-on times but the results do not depend
on them.)

Simulations are carried out using a 2-D code whose basic algorithm was
based on FARGO by \cite{Masset00} and \citet{RossbyIII}. The
co-rotating frame is used so that the positions of the central star
and the planet are fixed at $(r,\phi)=(0,0)$ and $(1,0)$
(acceleration due to frame rotation is also included, see
\citet{Kley98}).  Runs are made using several radial and azimuthal
grids to study the influence of resolution, and those presented here
have $(n_{r}\times n_{\phi})= 300\times 1200$.  Outflow boundary
conditions are used at both the inner and outer radial boundaries. The
simulations typically last several hundred orbits at $r=1$.

One key difference between our simulations and previous studies is
that our simulations are performed in the inviscid limit, i.e., we do
not explicitly include a viscosity term, though numerical viscosity is
inevitable and is needed to handle for example shocks.  In addition,
our simulations tend to be of higher resolution.  With a radial $n_r =
300$ (some runs go up to $n_r=600, n_\phi=2400$) and $0.4 \leq r \leq
2$, the diameter of the Hill sphere of a planet with $\mu=10^{-4}$ is
resolved by at least $\sim 12$ cells in each direction.  Using a
nested-grid technique, \citet{DAngeloHenningKley02} have also made
very high resolution simulations, though their simulations include an
explicit viscosity.

\section{Torque Evolution}

Our simulation results are generally in agreement with previous
simulations for the planet on a fixed circular orbit
\citep[see][]{Masset02} in terms of the generic structures produced in
the disk by the tidal interactions between the planet and its
surrounding flows.  Since the planet mass is relatively small, only
density dips are produced without gaps (at least during a few hundred
orbits simulated here).  We define the co-orbital region here as
$|\Delta r| =|r-1|\leq 6 r_H$.  The co-orbital region can be separated
into several parts depending on their streamline behavior
\citep[see][]{Masset02}, including the horseshoe (or librating) region
$|\Delta r| < r_H$, the separatrix region $r_H < |\Delta r| < \sqrt{12}
r_H$, and the streaming region $|\Delta r| > \sqrt{12} r_H$. The
horseshoe region can be further divided into real horseshoe
streamlines enclosing all three libration points ($L_{3,4,5}$) and
tadpole like streamlines enclosing either $L_4$ or $L_5$.

\begin{figure}
\epsscale{0.5}
\plotone{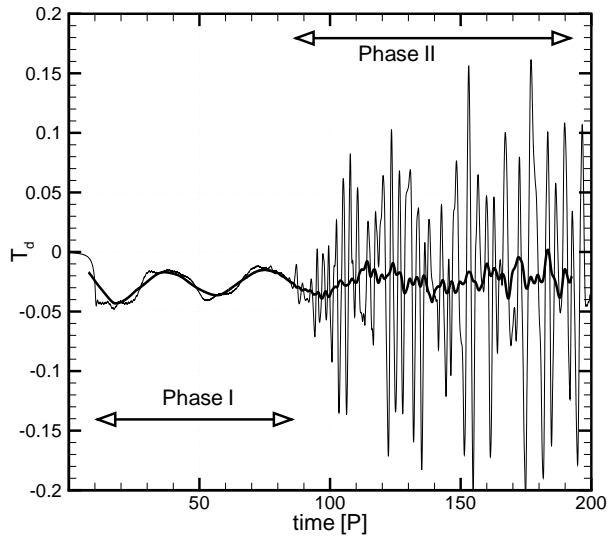}
\caption{Time evolution of the total torque $T_d$ exerted on the planet.
Phase~I ($t<90P$): The total torque is negative, smooth, and
shows an oscillation period of $\approx 40P$ which is consistent
with the libration period. We used a turn-on time of 10 orbits for 
the planet mass. Phase~II ($t>90P$): The torque starts to show
oscillations with increasingly high amplitudes.
During the transition from Phase~I to Phase~II, vortices emerge around
$\Delta r \approx 2.4 r_H$. These vortices experience close-encounters 
with the planet, resulting in large amplitude variations in the 
torque. The bold curve depicts a moving average of $T_d$ with a window
of $t=15P$ orbits.
\label{fig:torq_t}}
\end{figure}

Our key finding is shown in Fig. \ref{fig:torq_t} which displays the
total torque on the planet from the disk. It has two very distinct
phases: Phase I: the total torque is negative and smooth, modulated by
a period that is similar to the libration period near the L4 and L5
points ($\tau_{lib} \sim \sqrt{4/27\mu} \sim 40P$ where $P$ is one
orbit at $r=1$).  Phase II: the total torque shows very large
amplitude and fast oscillations with a quasi-period of a few orbits.
We see such a behavior in all the runs with three different
sound speeds, though the times when the ``phase'' transition occurs
are different. They are at $t\approx 40P, 90P,$ and $210P$ for $c_s =
0.04, 0.05$ and $0.06$, respectively.

\section{Shocks, Potential Vorticity Evolution and Instabilities}

\begin{figure}
\epsscale{1}
\plotone{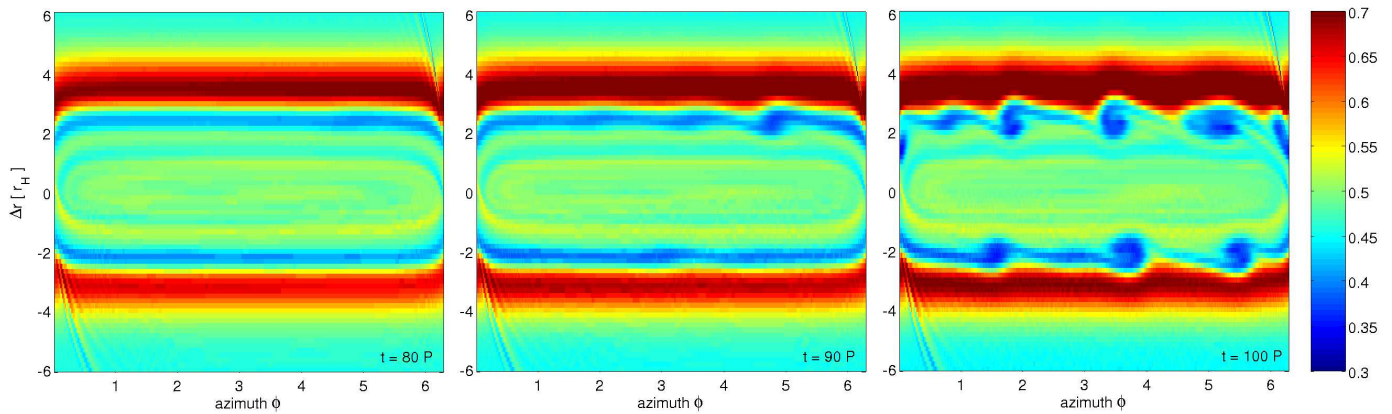}
\caption{Maps of potential vorticity $\zeta$ showing the production of
vortices. The planet is located at $(\Delta r,\phi)=(0,0)$
and the radial distance to the planet is given in units of
the Hill (Roche) radius $r_H$. Here, $c_s=0.05$. 
(left) At $t=80P$ the potential vorticity profile is relatively
smooth, with two ``channels'' at roughly $|\Delta r| \approx 2-3 r_H$.
(middle) At $t=90P$ the outer channel shows emerging vortices (anti-cyclones).
(right) At $t=100P$ vortices have grown to their full size in
both potential vorticity channels.  Note that vortices are moving
in the separatrix region, so they will experience
close-encounters with the planet. \label{fig:KH}}
\end{figure}

The obvious question is what causes such large amplitude and fast
changes in torques.  Closer inspection of the co-orbital region at
times around $90P$ (using $c_s=0.05$ as an example) reveals that
vortices are generated in the separatrix region.  This development is
shown in Fig. \ref{fig:KH}, which depicts maps of potential vorticity
$\zeta$ through the times when the system undergoes its ``phase''
transition.  One can see that these vortices are anti-cyclones which
are also blobs of higher densities (not shown here).  Being in the
separatrix region, they experience very close encounters with the
planet repeatedly, thus exerting strong ``impulses'' to the
planet. This results in the large amplitude and fast variations in the
torque evolution.

\begin{figure}
\epsscale{0.5}
\plotone{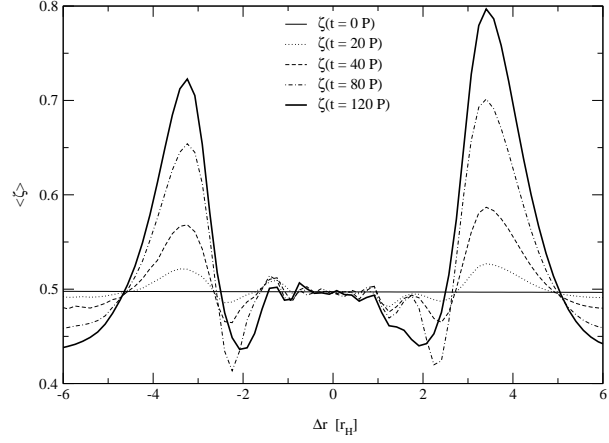}
\caption{Evolution of the azimuthally averaged potential vorticity
$\langle\zeta\rangle$ for $c_s=0.05$ at different times $t=0, 20P, 
40P, 80P, 120P$. Take the structure at $\Delta r \sim 2-3 r_H$ as
an example. The profile develops dips ($\Delta r=2.4r_H$) 
and peaks ($\Delta r=3.4 r_H$) progressively from $t = 0 - 80P$
until the vortices are produced at $t=90P$, after which
the minimum flattens (see the bold $t=120P$ curve).
\label{fig:pv}}
\end{figure}

To understand this phenomenon further, in Fig. \ref{fig:pv}, we plot
the azimuthally averaged radial potential vorticity profile
$\langle\zeta\rangle$ for $c_s=0.05$ at different times $t=0, 20P,
40P, 80P, 120P$. One sees that this profile deviates progressively
away from the initially flat radial profile, developing two prominent
structures at $|\Delta r| \approx 2-3 r_H$, with a minimum at
$\Delta r \approx 2.4r_H$ and a maximum at $\Delta r=3.4
r_H$. Comparing with Fig. \ref{fig:KH}, vortices apparently emerge
from the minima.

The existence of the extrema (or inflection points) in the potential
vorticity profile is usually regarded as a necessary condition for
non-axisymmetric instabilities in rotating shear flows (Drazin \& Reid
1981). The progressive steepening of this profile, say, at $\Delta r
\approx 2.4r_H$ until $t < 90P$ and the subsequent flattening (see the
curve at $t=120P$) strongly suggest that there is a secondary
instability which is excited around $\Delta r \approx 2.4r_H$ and
vortices are the non-linear outcome of such an instability.

How could the potential vorticity profile develop such inflection
points/structures from an initially flat and stable profile?
As can be seen from the evolution equation for potential vorticity
\begin{equation}
D(\zeta\hat{z})/Dt = (\zeta \hat{z} \cdot \nabla) {\bf v}
= 0~~{\rm in~2D},
\end{equation}
where $D/Dt$ is the Lagrangian derivative, that $\zeta$ is conserved
along the streamlines in 2D (Korycansky \& Papaloizou 1996). Thus, the
initial uniform $\zeta$ distribution should be preserved in an inviscid
disk.  This conclusion, however, is only true provided that
streamlines can be defined everywhere in a dissipationless flow.

Disk flows in the co-orbital region around the planet, however, do not
satisfy this condition (except perhaps for flows within $|\Delta r| <
r_H$) because of the two sets of spiral shocks produced by the
planet. These two sets of shocks emanate from close to the planet and
cut through the whole disk. Flow lines which encounter these shocks
are broken and the dissipation in shocks breaks the potential
vorticity conservation.  The consequence of the shock is equivalent to
that of a strong viscous stress in the momentum equation, the curl of
which, does not vanish.

\begin{figure}
\epsscale{0.5}
\plotone{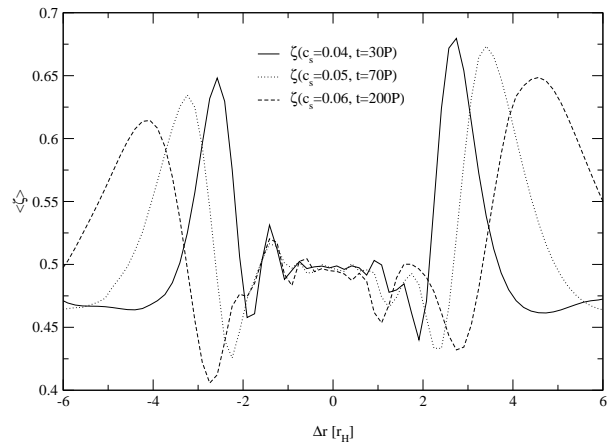}
\caption{Azimuthally averaged potential vorticity 
$\langle\zeta\rangle$ shortly before vortices set in for 
different sound speeds $c_s=0.04, 0.05, 0.06$. The
evolution for lower $c_s$ is much faster and therefore vortices
develop early on. Vortices first appear at the minima of the 
profile. The locations of the minima move outwards proportionally
as sound speeds increase, so do the starting locations of the
spiral shocks.  
\label{fig:pv3cs}}
\end{figure}

To establish the role of shocks in breaking the potential vorticity
conservation, we have performed three runs with different sound speeds
$c_s = 0.04, 0.05, 0.06$. Even though they only span a small range,
but the effects are quite clear. In Fig.  \ref{fig:pv3cs}, we plot the
azimuthally averaged potential vorticity profiles for these three runs
at a time just before the emergence of vortices. Vortices first appear
at the minimum of the ``valley'' in the profiles which moves outwards
as $c_s$ increases. This growth pattern is consistent with the fact
that the starting location of the shocks is also moving away from the
planet when the sound speed increases.

Note that the cause for the emergence and growth of inflection points
in the azimuthally averaged $\zeta$ distribution presented here is
different from that due to its re-arrangment along the horseshoe orbits
(Balmforth and Korycansky 2001).  In our calculation with a more
general initial $\Sigma$ (and $\zeta$) distribution, we confirm that
onset of the shearing instability occurs at an earlier epoch.  The
appearance of the subsequently self-excited vortices is quite similar
to the development of the Kelvin-Helmholtz (KH) instability observed
at shear interfaces. A detailed study on the properties of this
secondary instability, such as its threshold and growth rate, will be
presented in a forthcoming paper.

\section{Conclusions and Discussions}

In this paper, we consider the tidal interaction between a
protostellar disk and an embedded planet with a modest mass.  Linear
torque calculations suggest that such interaction leads to angular
momentum transfer and inward migration of the planets.  For planets
with mass comparable to that of the Earth, the inferred migration time
scale (Ward 1997) is much shorter than the observationally determined 
depletion time scale of the disk \citep{HaischLada01}.  This 
analytic result raises a problem for the formation of giant 
planets through the core accretion process (Pollack {\it et al.} 1996).  
We reinvestigate this process in an attempt to resolve this paradox.

The embedded protoplanet induces a circulation flow pattern near its
corotation region in the disk. We show here that even in
disks with initially uniform potential vorticities, inflection points
may emerge and grow spontaneously as potential vorticity is generated
near the spiral shock in the vicinity of the protoplanet.
These inflection points lead to the onset and growth of secondary
instabilities.

The existence of secondary instabilities and their non-linear outcome
as vortices indicate that the flows in the co-orbital region are
perhaps more complicated than linear analyses have suggested.  These
vortices should exert strong torques on the planet.

One limitation of our current study is that the planet is being
artificially held on a fixed circular orbit. Such strong and rapidly
varying torques from the vortices on the planet might cause the planet
to ``oscillate'' and lose its phase coherence with the surrounding
flow.  This feedback process should have important implications for
the Type I migration problem. Allowing the planet to migrate radially
must be included in order to understand the planet's response to such
torques. These issues will be addressed in a forthcoming publication.

\acknowledgments
We wish to thank Neil Balmforth, Darryl Holm and Edison Liang for useful
discussions. This research was performed under the auspices of
the Department of Energy. It was supported by the
Laboratory Directed Research and Development Program
at Los Alamos and by LANL/IGPP.  We are also supported in part by
NASA through NAG5-11779, NAG5-9223, and NSF through AST-9987417.



\clearpage


\end{document}